\documentclass{elsart5p}
\usepackage{graphics}
\usepackage{graphicx}
\usepackage{amsmath}
\usepackage{amssymb}
\begin{document}
\begin{frontmatter}
\title{Fermi Surface and Magnetism in the Kondo lattice: a Continuum Field Theory Approach }
\author{S. J. Yamamoto\corauthref{Yamamoto}},
\ead{seiji@rice.edu}
\author{Qimiao Si}
\address{Department of Physics and Astronomy, Rice University, Houston, Texas, 77251-1892}
\corauth[Yamamoto]{Corresponding author. Tel: (713) 348-4938 fax: (713) 348-4150}

\begin{abstract}
We consider the Fermi surface inside the antiferromagnetic
ordered
region of a Kondo lattice system in an arbitrary dimension higher
than one. We establish the existence of ${\rm AF_S}$, an 
antiferromagnetic phase whose Fermi surface is ``small,''
in the sense that the local
moments do not participate in the Fermi-surface
formation.
This is in contrast
to the ``large'' Fermi surface that is typically assumed for heavy 
fermion metals. 
We extend 
our earlier work 
to the case that
the Fermi surface of the conduction electrons 
intersects
the 
antiferromagnetic Brillouin zone boundary.
Our results provide a new perspective on local quantum criticality.
In addition, our results imply that, for the 
${\rm AF_S}$ phase, it is important to keep track 
of the
dynamical screening processes;
we suggest that this effect is not captured in a recent
variational Monte-Carlo study of the Kondo lattice.
\end{abstract}

\end{frontmatter}

Quantum criticality and the associated magnetic quantum phases of heavy
fermion metals are of extensive current interest, and the nature of the
Fermi surface has emerged as an important characterization
of both\cite{paschen2004, shishido2005}. This is because the presence
or absence of Kondo screening, which influences quantum criticality, 
also leads to distinct Fermi surfaces in the proximate zero-temperature
phases\cite{si2006}. In order to elucidate these issues, we study the
Kondo lattice
model deep in its antiferromagnetic (AF) phase,
in terms of a quantum nonlinear sigma model (QNL$\sigma$M) 
representation 
of the local-moment component\cite{yamamoto2007}.
We end up with
an effective coupling between the 
spin waves and conduction electrons,
which we show is exactly marginal in the renormalization
group (RG) sense.

The Kondo lattice model is defined by:
$\mathcal{H} = \mathcal{H}_c + \mathcal{H}_f +\mathcal{H}_K $,
where, in standard notation, 
$\mathcal{H}_c = \sum_{\vec{k}\sigma}
c^{\dagger}_{\vec{k}\sigma}
c_{\vec{k}\sigma}$ for a band of conduction
$c-$electrons, 
$\mathcal{H}_f = (1/2) \sum_{ij} I_{ij} {\vec S}_i \cdot {\vec S}_j$
for a lattice of spin-$1/2$ $f-$moments,
and 
$\mathcal{H}_K = \sum_{i} J_K {\vec S}_i \cdot {\vec s}_{c,i}$ 
describes the Kondo interaction.
We will focus
on the 
AF
Kondo lattice model,
with the dominant nearest-neighbor RKKY interaction $I$ being
antiferromagnetic ($I>0$), as is the Kondo interaction
($J_K>0)$. We consider the situation deep inside the 
AF part of the 
$T=0$ phase diagram, where $J_K \ll I$ and both are much smaller
than the 
bandwidth of the conduction electrons.

Since we are dealing with the ordered phase of a local moment 
antiferromagnet
rather than a single impurity spin, 
we 
map
$\mathcal{H}_f$ to a QNL$\sigma$M
by standard means\cite{haldane1983, chakravarty1989}:
$2 \vec{S}_i 
\to \eta_{\vec{x}} \vec{n}(\vec{x},\tau)\sqrt{1-
[2a^d \vec{L}(\vec{x},\tau) 
]^2} 
+ 
2a^d \vec{L}(\vec{x},\tau) $,
where $\vec{x}$ labels the position,
$\eta_{\vec{x}} = \pm 1$ on even and odd sites,
and $a$ is the lattice constant. 
The low-lying excitations are concentrated in the momentum space near
${\vec q}={\vec Q}$ (the AF wavevector) 
and ${\vec q}={\vec 0}$, corresponding to 
$\vec{n}$ and $\vec{L}$ respectively. 

The case with the Fermi surface of the conduction electrons not
intersecting the antiferromagnetic Brillouin zone (AFBZ) boundary
was treated in Ref.~\cite{yamamoto2007}.
Here, the linear coupling $\vec{n}\cdot\vec{s}_c $ cannot
connect
two points on the Fermi surface and thus, for low energy
physics, it does not come into play.
The Kondo coupling reduces to,
$\vec{S} \cdot\vec{s}_c \rightarrow a^d \vec{L}\cdot\vec{s}_c$,
corresponding to forward scattering for the conduction
electrons.
Integrating out 
the $\vec{L}$ field leads to the effective action:
$\mathcal{S} = \mathcal{S}_{\text{QNL}\sigma\text{M}}
+\mathcal{S}_{\text{Berry}}+\mathcal{S}_K+\mathcal{S}_c$,
where 
$\mathcal{S}_{\text{QNL}\sigma\text{M}}$ is the quantum non-linear
sigma model and $S_c$ the action for a free fermion band
with a dispersion of $\xi_K \equiv v_F(K-K_F)$. The Kondo coupling 
has the form
$\mathcal{S}_K = \lambda\int d^dx d\tau\left[ \vec{s}_c(\vec{x},\tau)
\cdot \vec{\varphi}(\vec{x},\tau) \right]$,
where the vector boson field $\vec{\varphi}$ 
represents  $\vec{n}\times\frac{\partial\vec{n}}{\partial\tau}$,
with $\vec{n}$ being the QNL$\sigma$M field.
The constraint $\vec{n}^2=1$ is implemented by
$\vec{n} = (\vec{\pi}, \sigma)$, where 
$\vec{\pi}$ labels the Goldstone magnons
and $\sigma \equiv \sqrt{1-\vec{\pi}^2}$ is the massive field.
The Berry phase term,
$\mathcal{S}_{\text{Berry}}$, 
is not important 
inside the N\'{e}el phase.

In order to keep track of the Fermi surface in the RG
procedure, we use a combination of the fermionic RG~\cite{shankar1994}
and standard bosonic RG methods.
We find a marginal coupling at the tree level:
the scaling dimension
$[\lambda] = 0$.  
It turns out that certain kinematic restrictions
prevent 
higher-loop corrections
from entering the beta function.  
This results from the forward-scattering nature of the effective Kondo
coupling. At the one-loop level this can be seen by an explicit 
calculation, where momentum conservation inside the restrictive
cutoffs limits the region of integration to an area of size
$(d\Lambda)^{3/2}$.
(Here $d\Lambda = \Lambda - \Lambda/s$, in the $s \rightarrow 1^+$ limit.)
Going beyond one loop,
it turns out that the Kondo vertex contains
a small parameter, $1/N_{\Lambda} \equiv \Lambda/K_F$, relative
to the kinetic term: 
$\mathcal{S}_K/\mathcal{S}_c \propto 1/\sqrt{N_{\Lambda}}$ in the
spin-flip case, and 
$\mathcal{S}_K/\mathcal{S}_c \propto 1/N_{\Lambda}$ in the 
longitudinal case.
In the asymptotic 
low-energy limit ($N_{\Lambda} = \infty$),
the higher loop corrections to the beta function vanish.
Therefore, 
the one-loop result is the whole story and the Kondo coupling 
is exactly marginal. 
There is no flow to the strong-coupling 
fixed point.
This implies the absence of 
static Kondo screening, and the 
Fermi surface remains small.
It also means that there is no singular 
correction to the QNL$\sigma$M itself.
Both conclusions can also be seen in a suitable large-N limit
of the effective action\cite{yamamoto2007}.

\begin{figure}[tp]
   \centering
   \includegraphics[width=0.20\textwidth]{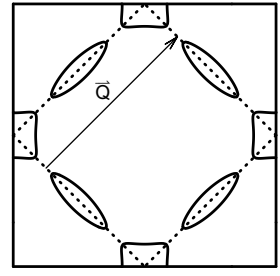}
   \caption{
When the Fermi surface of the conduction electrons
intersects the AFBZ
boundary (the dashed lines),
the Kondo coupling connects the re-diagonalized
fermions (whose Fermi surfaces are given by the solid lines)
to the QNL$\sigma $M fields. 
}
\label{fig:gappedFS}
\end{figure}

We now turn to the case when the Fermi surface of the conduction electrons
intersects the AFBZ boundary. Here, the linear coupling 
$\vec{n}\cdot\vec{s}_c $ cannot be neglected. The AF order
of the local moments implies that a staggered field is applied
to the conduction electrons, resulting in a reconstruction of their Fermi
surface: the hot spots of the Fermi surface become gapped out,
as shown in Fig.~\ref{fig:gappedFS}. At the mean field level,
the conduction electron component now becomes:
\begin{eqnarray}
	\mathcal{H}_{c}^{\text{MF}} 
	&=&  \sum_{k\alpha\beta}^{AFBZ} \left(\begin{array}{cc} c^{\dagger}_{k,\alpha}, & c^{\dagger}_{k+Q,\alpha}\end{array} \right)
	\left(
	\begin{array}{cc}
		\tau^0_{\alpha\beta}\epsilon_k & \tau^z_{\alpha\beta}\Delta \\
		\tau^z_{\alpha\beta}\Delta & \tau^0_{\alpha\beta}\epsilon_{k+Q}
	\end{array}
	\right)
	\left( \begin{array}{c} c_{k,\beta} \\ c_{k+Q,\beta} \end{array}\right) \nonumber
\end{eqnarray}
where 
the sum on $k$ only runs over the AFBZ,
$\vec{Q}=(\pi/a,\pi/a)$ is the AF ordering wavevector, 
$\tau^{0,z}$ are the $2\times 2$ unit/Pauli matrices, 
and the gap is given by the product of the Kondo coupling
and expectation value of the massive field of the QNL$\sigma$M:
$\Delta = \lambda\langle \sigma \rangle$.  
This is simply diagonalized by a unitary transformation:
\begin{eqnarray}
	\left( \begin{array}{c} a_{k\alpha} \\ b_{k\alpha} \end{array}\right)
	&=&
	\left( \begin{array}{cc}
		u_k\sigma^0_{\alpha\beta} &  v_k \sigma^z_{\alpha\beta} \\
		v_k\sigma^z_{\alpha\beta} & -u_k \sigma^0_{\alpha\beta}
		\end{array}\right)
	\left( \begin{array}{c} c_{k\beta} \\ c_{k+Q,\beta} \end{array}\right) .
\end{eqnarray}
Using these new quasiparticles, the effective spin-flip Kondo couplings
become 
\begin{eqnarray}
&&-\frac{J_K}{2}
	 \sum_{k}^{AFBZ} \sum_q^{BZ}\Big[ 
	\Gamma^{\sigma\bar{\sigma}}(k,q)\Big(
	 a^{\dagger}_{k\sigma}a_{k+q,\bar{\sigma}} - b^{\dagger}_{k\sigma}b_{k+q,\bar{\sigma}}
	\Big) n^{\bar{\sigma}}_q 
	\Big]
\end{eqnarray}
where $\Gamma^{\sigma\bar{\sigma}}(k,q) = u_k v_{k+q} - v_k u_{k+q}$ is the
coherence factor. The
other terms, such as inter-band interactions (e.g. $a^{\dagger} b$)
are gapped out at low energies.  
Near the ordering wavevector the vertex is linear in momentum:
$\Gamma^{\uparrow\downarrow}(k,q) \propto q$, where $\vec{q}$
is the deviation from the AF ordering wavevector 
$\vec{Q}$.
(See {\it e.g.}, Ref.~\cite{vekhter2004}.)
The form of this linear-momentum suppression factor survives
beyond
the mean-field treatment of the conduction electron band,
as dictated by Adler's Theorem.
Within the RG analysis,
the linear-momentum factor
serves the same function as the time derivative in $\vec{\varphi}$
to preserve the marginality of the transverse Kondo coupling.

The marginal nature of the Kondo coupling means that
the effective Kondo coupling 
is finite at finite energies
in the ${\rm AF_S}$ phase. 
One corollary is that the ground state wavefunction
will incorporate such dynamical screening effects.
Recently, Watanabe and Ogata~\cite{watanabe07} have carried
out a variational Monte-Carlo study of the Kondo lattice.
Their choice of the variational wavefunction for the
${\rm AF_S}$ does not contain any dynamical screening,
which we believe is responsible for their finding 
that the ${\rm AF_S}$ phase is energetically
unfavorable.

In conclusion, we have shown that the Kondo coupling deep inside
the ordered region of the antiferromagnetic Kondo lattice is exactly
marginal,
thereby establishing the existence of an antiferromagnet with a 
small Fermi surface.
The stability of this ${\rm AF_S}$ phase provides a new anchoring point
to view the destruction of Kondo effect at the magnetic quantum critical
point, as given in the local quantum criticality~\cite{si2001}.
It also serves as a benchmark for any approximate or numerical studies
of the Kondo lattice model.

This work has been supported in part by
NSF, the Robert A. Welch Foundation and the W. M. Keck Foundation.

\end{document}